\documentclass[showkeys,superscriptaddress,floatfix]{revtex4-2}
\usepackage{graphicx,epstopdf}
\usepackage[caption=false]{subfig}
\usepackage{appendix}
\usepackage[T1]{fontenc}
\usepackage{lmodern}
\usepackage[dvipsnames,x11names]{xcolor}
\usepackage[colorlinks=true,linkcolor=ForestGreen,citecolor=ForestGreen,urlcolor=NavyBlue]{hyperref}
\usepackage[sort&compress]{natbib}
\usepackage{lipsum}
\usepackage{morefloats}
\usepackage[pdf]{pstricks}
\usepackage{amsmath}
\usepackage{amssymb}
\usepackage{amsfonts}
\usepackage{rotating}
\usepackage{cancel}
\usepackage{mathtools}
\usepackage{bbm}
\usepackage{dsfont}
\usepackage{bbold}
\usepackage{multirow}
\usepackage{ulem}
\usepackage{physics}
\usepackage{orcidlink}
\usepackage{colortbl}

\def\xslash{x\!\!\!\slash }

\def\vel{\left|}
\def\ver{\right|}

\begin{document}

\title{Study on the electromagnetic properties of the $[sc] [\bar q \bar b]$ and $[sc] [\bar s \bar b]$ states with $J^P = 1^+$ }

\author{Ula\c{s}~\"{O}zdem\orcidlink{0000-0002-1907-2894}}%
\email[]{ulasozdem@aydin.edu.tr }
\affiliation{ Health Services Vocational School of Higher Education, Istanbul Aydin University, Sefakoy-Kucukcekmece, 34295 Istanbul, T\"{u}rkiye}

 
\begin{abstract}
A systematic study of the electromagnetic properties of exotic states is conducted to elucidate their nature, which continues to be the subject of controversy and incomplete understanding in the field. In this study, the magnetic dipole and quadrupole moments of the tetraquarks $[sc] [\bar q \bar b]$ and $[sc] [\bar s \bar b]$ with spin-parity quantum numbers $J^P = 1^+$ are extracted in the context of a compact diquark-antidiquark configuration with the help of the QCD light-cone rules method.  The magnetic dipole moments are given  as $\mu_{[sc] [\bar u \bar b]} = -2.12^{+0.74}_{-0.59} \mu_N$, $\mu_{[sc] [\bar d \bar b]} = 1.66^{+0.60}_{-0.46} \mu_N$, and  $\mu_{[sc] [\bar s \bar b]} =  2.01^{+0.61}_{-0.50} \mu_N$.  The order of magnitude of the magnetic dipole moments would suggest that these outcomes may be achievable in forthcoming experiments. The magnetic and quadrupole moments of hadrons represent another important observable, along with their mass and decay width, which contribute to our understanding of the underlying quark structure and dynamics. Therefore, we hope that the results of this study will prove useful in theoretical and experimental investigations, which we anticipate will be an interesting research topic.
\end{abstract}

\maketitle

\section{motivation}\label{motivation}

While the idea of hadrons with more intricate structures than mesons with $q\bar q$ and baryons with $qqq$ has been known for a long time, the confirmation of the existence of the exotic state designated X(3872) employing experimental evidence was achieved by the Belle Collaboration in 2003~\cite{Belle:2003nnu}.  Since that time, numerous exotic hadron candidates have been observed by different collaborations. With each new observation, the family of such exotic states is expanding and enriching. The experimental discovery of these exotic states represents a dynamic and vibrant field within hadron physics, encompassing both experimental and theoretical methodologies. The most recent developments in the field of exotic states, as well as experimental and theoretical works, are listed in Refs.~\cite{Esposito:2014rxa,Esposito:2016noz,Olsen:2017bmm,Lebed:2016hpi,Nielsen:2009uh,Brambilla:2019esw,Agaev:2020zad,Chen:2016qju,Ali:2017jda,Guo:2017jvc,Liu:2019zoy,Yang:2020atz,Dong:2021juy,Dong:2021bvy,Meng:2022ozq, Chen:2022asf}. Nevertheless, it is crucial to acknowledge that despite the extensive theoretical and experimental investigations conducted, the internal structure, decay characteristics, and quantum numbers of these exotic states remain among the pivotal questions that remain unanswered. 

A substantial majority of the experimental explorations of exotic states to date have yielded either hidden-charm/bottom or doubly-charm outcomes. Another class of exotic states, tetraquarks involving heavy $c\bar b$ diquarks, is likely to exist.  It is important to note that these states have not been discovered experimentally. However, there are theoretical studies in the literature that investigate their existence and study their properties~\cite{Zhang:2009vs,Zhang:2009em,Sun:2012sy,Albuquerque:2012rq,Chen:2013aba, Agaev:2016dsg,Agaev:2017uky,Wang:2020jgb,Wang:2019xzt,Wu:2018xdi,Ortega:2020uvc}. One of the primary reasons for this phenomenon is that tetraquarks constructed from $c\bar b$ may exhibit stability against strong and electromagnetic decays. Due to these properties, they are highly valuable for the study of heavy-quark dynamics and the understanding of the dynamics of the QCD at a deeper level.  Further research is required from a range of theoretical viewpoints to gain comprehensive insight into the properties of tetraquarks with $c \bar b$, which will help future experimental studies. In addition to the mass and decay properties of these states, it is crucial to elucidate the substructure and geometric shape of tetraquarks with $c \bar b$. It would be advantageous to examine their electromagnetic and radiative transition properties to gain a more profound comprehension of their intrinsic nature. The literature contains several studies on the electromagnetic properties of tetraquarks, which are designed to elucidate their internal structure~\cite{Ozdem:2024dbq,Ozdem:2024lpk,Mutuk:2023oyz,Wang:2023bek,Ozdem:2023rkx,Ozdem:2023frj,Lei:2023ttd,Zhang:2021yul,Azizi:2023gzv,Ozdem:2022eds,Ozdem:2022kck,Xu:2020qtg,Wang:2017dce,Ozdem:2022yhi,Wang:2023vtx,Ozdem:2021hmk,Azizi:2021aib,Ozdem:2021hka,Xu:2020evn,Ozdem:2021yvo,Ozdem:2017exj,Ozdem:2017jqh}.

The purpose of this study is to systematically extract magnetic dipole and quadrupole moments of the $[sc] [\bar q \bar b]$ and $[sc] [\bar s \bar b]$ tetraquarks. To this end, a compact diquark-antiquark interpolating current with $J^P = 1^+$ spin-parity quantum numbers is constructed. Magnetic dipole and quadrupole moments are parameters of the nonperturbative domain of QCD. Therefore, to extract these parameters a reliable method, such as the QCD light cone sum rules (LCSR) method, is required. The LCSR method yields highly efficacious and predictable outcomes and represents a robust non-perturbative approach for investigating the dynamic and static attributes of conventional and exotic hadrons. The LCSR method entails calculating the correlation function in two distinct representations: one about hadrons (referred to as the "hadronic part") and the other to quark-gluon degrees of freedom (referred to as the "QCD part"). These two representations are then matched and sum rules are derived for the respective physical quantities.

This paper is organized as follows.  After Introduction, the formalism of QCD light-cone sum rules for the multipole moments of $[sc] [\bar q \bar b]$ and $[sc] [\bar s \bar b]$ tetraquarks with $J^P = 1^+$ are briefly introduced in Sec. \ref{formalism}. We present the numerical results and discussions for these states in Sec. \ref{numerical}. The final section is dedicated to our concluding remarks.

 \begin{widetext}
 
\section{Deriving the QCD light-cone sum rules for multipole moments}\label{formalism}

The multipole moments of the $[sc] [\bar q \bar b]$ and $[sc] [\bar s \bar b]$ ($Z_{c \bar b}$ for short) tetraquarks are obtained by the LCSR technique through the construction of a correlation function expressed as follows: 

\begin{equation}
 \label{edmn01}
\Pi _{\alpha \beta }(p,q)=i\int d^{4}x\,e^{ip\cdot x}\langle 0|\mathcal{T}\{J^i_{\alpha}(x)
J_{\beta }^{i \dagger }(0)\}|0\rangle_{\gamma}, 
\end{equation}
where $J^i_{\alpha(\beta)}(x)$ are the interpolating currents and  the subindex $\gamma$ represents the external electromagnetic field. The interpolating currents for the considered states are given by $J^i_{\alpha(\beta)}(x)$ with quantum numbers $J^{P} = 1^{+}$, which are expressed as follows:
\begin{align}
J_{\alpha }^{1}(x) &= \big[s^{a^T} (x) C \gamma_5 c^b (x)\big]\big[\bar q^a (x) \gamma_\alpha C \bar b^{b^T} (x) + \bar q^b (x) \gamma_\alpha C \bar b^{a^T} (x)\big],\\
J_{\alpha }^{2}(x) &= \big[s^{a^T} (x) C \gamma_5 c^b (x)\big]\big[\bar s^a (x) \gamma_\alpha C \bar b^{b^T} (x) + \bar s^b (x) \gamma_\alpha C \bar b^{a^T} (x)\big],
\label{curr}
\end{align}
where $a$ and $b$ are color indices, $C$ is the charge conjugation operator and; 
$q(x)$ denotes the $u(x)$ or  $d(x)$. 


 At the hadronic level, we plug into the correlation function a complete set of intermediate hadronic states with the same quantum numbers as the interpolating currents.  Inserting the intermediate states, and isolating the ground state contributions yield the following result:
\begin{align}
\label{edmn04}
\Pi_{\alpha\beta}^{Had} (p,q) &= {\frac{\langle 0 \mid J_\alpha (x) \mid
Z_{c \bar b}(p) \rangle}{p^2 - m_{Z_{c \bar b}}^2}} \langle Z_{c \bar b} (p) \mid Z_{c \bar b} (p+q) \rangle_\gamma 
\frac{\langle Z_{c \bar b} (p+q) \mid J_{\beta }^{\dagger } (0) \mid 0 \rangle}{(p+q)^2 - m_{Z_{c \bar b}}^2} + \mbox{higher states},
\end{align}
The $\langle 0 \mid J_\alpha (x) \mid Z_{c \bar b}(p) \rangle$ in Eq. (\ref{edmn04}) is expressed regarding hadron characters, as indicated below
\begin{align}
\label{edmn05}
\langle 0 \mid J_\alpha (x) \mid Z_{c \bar b} (p) \rangle =  \lambda_{Z_{c \bar b}} \varepsilon_\alpha^\theta\,,
\end{align}
with  $\lambda_{Z_{c \bar b}}$ and $ \varepsilon_\alpha^\theta\ $   are the current coupling and the polarization vector of particles under question, respectively.  
In Eq. (\ref{edmn04}), there is another matrix element, known as the radiative transition matrix element, which is expressed as~\cite{Brodsky:1992px}
\begin{align}
\label{edmn06}
\langle Z_{c \bar b}(p,\varepsilon^\theta) \mid  Z_{c \bar b} (p+q,\varepsilon^{\delta})\rangle_\gamma &= - \varepsilon^\tau (\varepsilon^{\theta})^\mu (\varepsilon^{\delta})^\nu
\Bigg[ G_1(Q^2)~ (2p+q)_\tau ~g_{\mu\nu}  
+ G_2(Q^2)~ ( g_{\tau\nu}~ q_\mu -  g_{\tau\mu}~ q_\nu)
\nonumber\\ 
&
- \frac{1}{2 m_{Z_{c \bar b}}^2} G_3(Q^2)~ (2p+q)_\tau 
q_\mu q_\nu  \Bigg],
\end{align}
where $G_i(Q^2)$'s being form factors of the corresponding transition, with  $Q^2=-q^2$.   
The correlation function is derived from Eqs.~(\ref{edmn04})-(\ref{edmn06}) in the following manner: 
%
\begin{align}
\label{edmn09}
 \Pi_{\alpha\beta}^{Had}(p,q) &=  \frac{\varepsilon_\rho \, \lambda_{Z_{c \bar b}}^2}{ [m_{Z_{c \bar b}}^2 - (p+q)^2][m_{Z_{c \bar b}}^2 - p^2]}
 \bigg\{G_1(Q^2)(2p+q)_\rho\bigg[g_{\alpha\beta}-\frac{p_\alpha p_\beta}{m_{Z_{c \bar b}}^2}
 -\frac{(p+q)_\alpha (p+q)_\beta}{m_{Z_{c \bar b}}^2}+\frac{(p+q)_\alpha p_\beta}{2m_{Z_{c \bar b}}^4}\nonumber\\
 & \times (Q^2+2m_{Z_{c \bar b}}^2)
 \bigg]
 + G_2 (Q^2) \bigg[q_\alpha g_{\rho\beta}  
 - q_\beta g_{\rho\alpha}-
\frac{p_\beta}{m_{Z_{c \bar b}}^2}  \big(q_\alpha p_\rho - \frac{1}{2}
Q^2 g_{\alpha\rho}\big) 
+
\frac{(p+q)_\alpha}{m_{Z_{c \bar b}}^2}  \big(q_\beta (p+q)_\rho+ \frac{1}{2}
Q^2 g_{\beta\rho}\big) 
\nonumber\\
&-  
\frac{(p+q)_\alpha p_\beta p_\rho}{m_{Z_{c \bar b}}^4} \, Q^2
\bigg]
-\frac{G_3(Q^2)}{m_{Z_{c \bar b}}^2}(2p+q)_\rho \bigg[
q_\alpha q_\beta -\frac{p_\alpha q_\beta}{2 m_{Z_{c \bar b}}^2} Q^2 
+\frac{(p+q)_\alpha q_\beta}{2 m_{Z_{c \bar b}}^2} Q^2
-\frac{(p+q)_\alpha q_\beta}{4 m_{Z_{c \bar b}}^4} Q^4\bigg]
\bigg\}\,.
\end{align}
The magnetic ($F_M(Q^2)$) and quadrupole ($F_{\mathcal D}(Q^2)$) form factors can be derived in the form of the previously defined form factors, $G_i(Q^2)$, as follows: 
\begin{align}
\label{edmn07}
&F_M(Q^2) = G_2(Q^2)\,,\nonumber \\
&F_{\mathcal D}(Q^2) = G_1(Q^2)-G_2(Q^2)+\Big(1+\frac{Q^2}{4 m_{\mathrm{Z_{c \bar b}}}^2}\Big) G_3(Q^2)\,.
\end{align}
In the static limit ($Q^2=0$), where we are dealing with the real photon, the $F_M(0)$ and $F_{\mathcal D}(0)$ can be described as the magnetic moment ($\mu$) and quadrupole $(\mathcal D)$ moments by the following relationship: 
\begin{align}
\label{edmn08}
 \mu  &= \frac{e}{2 m_{Z_{c \bar b}}}\,F_M(0), 
 \\
\mathcal D  &= \frac{e}{ m^2_{Z_{c \bar b}}}\,F_{\mathcal D}(0).
\end{align}
The results of the analysis at the hadron level are extracted in the form of a representation of the physical parameters under study. To obtain a representation of the analysis at the quark-gluon level, it is necessary to proceed with further calculations. 

At the quark-gluon level, the correlation function can be evaluated by the operator product expansion (OPE) technique and expressed in terms of photon distribution amplitudes. 
 Upon completion of elementary mathematical operations, the following outcome for the $Z_{c \bar b}$ states is derived: 
\begin{align}
\Pi _{\alpha \beta }^{\mathrm{QCD}-J_\alpha^1}(p,q)&=i\int d^{4}xe^{ip\cdot x} \langle 0 \mid \bigg\{  \mathrm{Tr}%
\Big[ \gamma _{\alpha }\widetilde{S}_{b}^{b^{\prime }b}(-x)\gamma _{\beta
}   
S_{q}^{a^{\prime }a}(-x)\Big]    \mathrm{Tr}\Big[ \gamma _{5} \widetilde{S}_{q}^{aa^{\prime
}}(x)\gamma _{5}S_{c}^{bb^{\prime }}(x)\Big] \nonumber\\
&+\mathrm{Tr}\Big[ \gamma
_{\alpha }\widetilde{S}_{b}^{a^{\prime }b}(-x) \gamma _{\beta }S_{q}^{b^{\prime }a}(-x)\Big]
\mathrm{Tr}%
\Big[ \gamma _{5}\widetilde{S}_{q}^{aa^{\prime }}(x)\gamma
_{5}S_{c}^{bb^{\prime }}(x)\Big]  \nonumber \\
&+\mathrm{Tr}\Big[ \gamma _{\alpha }\widetilde{S}_{b}^{b^{\prime
}a}(-x)\gamma _{\beta }S_{q}^{a^{\prime }b}(-x)\Big] 
\mathrm{Tr}\Big[
\gamma _{5}\widetilde{S}_{q}^{aa^{\prime }}(x)\gamma _{5}S_{c}^{bb^{\prime
}}(x)\Big]  \notag \\
& +\mathrm{Tr}\Big[ \gamma _{\alpha }\widetilde{S}_{b}^{a^{\prime
}a}(-x)\gamma _{\beta }S_{q}^{b^{\prime }b}(-x)\Big]
\mathrm{Tr}\Big[
\gamma _{5}\widetilde{S}_{q}^{aa^{\prime }}(x) \gamma _{5}S_{c}^{bb^{\prime
}}(x)\Big] \bigg\} \mid 0 \rangle_{\gamma} ,  \label{eq:QCDSide}
\end{align}%
\begin{align}
\Pi _{\alpha \beta }^{\mathrm{QCD}-J_\alpha^2}(p,q)&=i\int d^{4}xe^{ip\cdot x} \langle 0 \mid \bigg\{  \mathrm{Tr}%
\Big[ \gamma _{\alpha }\widetilde{S}_{b}^{b^{\prime }b}(-x)\gamma _{\beta
}   
S_{s}^{a^{\prime }a}(-x)\Big]    \mathrm{Tr}\Big[ \gamma _{5} \widetilde{S}_{s}^{aa^{\prime
}}(x)\gamma _{5}S_{c}^{bb^{\prime }}(x)\Big] \nonumber\\
&+\mathrm{Tr}\Big[ \gamma
_{\alpha }\widetilde{S}_{b}^{a^{\prime }b}(-x) \gamma _{\beta }S_{s}^{b^{\prime }a}(-x)\Big]
\mathrm{Tr}%
\Big[ \gamma _{5}\widetilde{S}_{s}^{aa^{\prime }}(x)\gamma
_{5}S_{c}^{bb^{\prime }}(x)\Big]  \nonumber \\
&+\mathrm{Tr}\Big[ \gamma _{\alpha }\widetilde{S}_{b}^{b^{\prime
}a}(-x)\gamma _{\beta }S_{s}^{a^{\prime }b}(-x)\Big] 
\mathrm{Tr}\Big[
\gamma _{5}\widetilde{S}_{s}^{aa^{\prime }}(x)\gamma _{5}S_{c}^{bb^{\prime
}}(x)\Big]  \notag \\
& +\mathrm{Tr}\Big[ \gamma _{\alpha }\widetilde{S}_{b}^{a^{\prime
}a}(-x)\gamma _{\beta }S_{s}^{b^{\prime }b}(-x)\Big]
\mathrm{Tr}\Big[
\gamma _{5}\widetilde{S}_{s}^{aa^{\prime }}(x) \gamma _{5}S_{c}^{bb^{\prime
}}(x)\Big] \bigg\} \mid 0 \rangle_{\gamma} ,  \label{eq:QCDSide2}
\end{align}%
where the $S_{Q}(x)$ and $S_{q}(x)$ are the propagators of heavy and light quarks with $\widetilde{S}_{Q(q)}^{ij}(x)=CS_{Q(q)}^{ij\rm{T}}(x)C$.  The explicit forms of these propagators are defined as follows:~\cite{Yang:1993bp, Belyaev:1985wza},
\begin{align}
\label{edmn13}
S_{q}(x)&= S_q^{free}(x) 
- \frac{\langle \bar qq \rangle }{12} \Big(1-i\frac{m_{q} \xslash}{4}   \Big)
- \frac{ \langle \bar qq \rangle }{192}
m_0^2 x^2  \Big(1 
  -i\frac{m_{q} \xslash}{6}   \Big)
+\frac {i g_s~G^{\alpha \beta} (x)}{32 \pi^2 x^2} 
\bigg[\rlap/{x} 
\sigma_{\alpha \beta} +  \sigma_{\alpha \beta} \rlap/{x}
 \bigg],\\
%
S_{Q}(x)&=S_Q^{free}(x)
-\frac{m_{Q}\,g_{s}\, G^{\alpha \beta}(x)}{32\pi ^{2}} \bigg[ (\sigma _{\alpha \beta }{\xslash}
+{\xslash}\sigma _{\alpha \beta }) 
    \frac{K_{1}\big( m_{Q}\sqrt{-x^{2}}\big) }{\sqrt{-x^{2}}}
 +2\sigma_{\alpha \beta }K_{0}\big( m_{Q}\sqrt{-x^{2}}\big)\bigg],
 \label{edmn14}
\end{align}%
with  
\begin{align}
 S_q^{free}(x)&=\frac{1}{2 \pi x^2}\Big(i \frac{\xslash}{x^2}- \frac{m_q}{2}\Big),\\
 S_Q^{free}(x)&=\frac{m_{Q}^{2}}{4 \pi^{2}} \bigg[ \frac{K_{1}\big(m_{Q}\sqrt{-x^{2}}\big) }{\sqrt{-x^{2}}}
+i\frac{{\xslash}~K_{2}\big( m_{Q}\sqrt{-x^{2}}\big)}
{(\sqrt{-x^{2}})^{2}}\bigg],
\end{align}
where $m_0$ being the quark-gluon mixed condensate with $ m_0^2= \langle 0 \mid \bar  q\, g_s\, \sigma_{\alpha\beta}\, G^{\alpha\beta}\, q \mid 0 \rangle / \langle \bar qq \rangle $, $G^{\alpha\beta}$ is the gluon field-strength tensor, and $K_i$'s are the modified Bessel functions of the second type.  

To perform further calculations at the quark-gluon level, it is necessary to consider two distinct contributions from the short- and long-distance interactions of the photon with quarks.   To ascertain the nature of the short-distance interactions, it is necessary to proceed with the subsequent replacement as described below: 
\begin{align}
\label{free}
S^{free}(x) \longrightarrow \int d^4z\, S^{free} (x-z)\,\rlap/{\!A}(z)\, S^{free} (z)\,.
\end{align}
To include long-distance interactions in the analysis, it is convenient to apply the formula below:
 \begin{align}
\label{edmn21}
S_{\alpha\beta}^{ab}(x) \longrightarrow -\frac{1}{4} \big[\bar{q}^a(x) \Gamma_i q^b(0)\big]\big(\Gamma_i\big)_{\alpha\beta},
\end{align}
where $\Gamma_i = \{\textbf{1}$, $\gamma_5$, $\gamma_\alpha$, $i\gamma_5 \gamma_\alpha$, $\sigma_{\alpha\beta}/2\}$. In calculating short- and long-distance contributions, we employ a single propagator in the equation, whereas the remaining propagators are treated as full propagators.  Upon the inclusion of long-distance contributions in the analysis, matrix elements such as $\langle \gamma(q)\vel \bar{q}(x) \Gamma_i G_{\alpha\beta}q(0) \ver 0\rangle$ and $\langle \gamma(q)\vel \bar{q}(x) \Gamma_i q(0) \ver 0\rangle$ emerge. The aforementioned matrix elements, expressed in the form of photon wave functions, are crucial parameters in the calculation of long-distance interactions (see Ref.~\cite{Ball:2002ps} for details on photon distribution amplitudes (DAs)). It is important to note that the photon DAs utilized in this study include contributions solely from light quarks. However, in principle, a photon can be emitted long-distance from heavy quarks. However, such long-distance photon emission from heavy quarks is highly suppressed owing to the large mass of the heavy quarks. This kind of contribution is neglected in our analysis. As explained in Eq.~(\ref{free}), only the short-distance photon emission from heavy quarks is considered. For this reason, DAs including heavy quarks are not considered in our analysis. The aforementioned procedures have yielded the quark-gluon correlation function, which has been obtained through the use of quark-gluon parameters and photon DAs. 

The above-mentioned procedures give the sum rules for the magnetic and quadrupole moments of the $[sc] [\bar q \bar b]$ and $[sc] [\bar s \bar b]$ tetraquarks as follows, 
\begin{align}
\label{jmu1}
 \mu_{Z_{c \bar b}}^{J_\alpha^1}\,  &= \frac{e^{\frac{m_{Z_{c \bar b}}^{2}}{\rm{M^2}}}}{ \lambda^{2}_{Z_{c \bar b}}} \,\, \rho_1(\rm{M^2},\rm{s_0}),~~~~
 \mathcal{D}_{\mathrm{Z_{c \bar b}}}^{J_\alpha^1} = m_{\mathrm{Z_{c \bar b}}}^2 \frac{e^{\frac{m_{Z_{c \bar b}}^{2}}{\rm{M^2}}}}{ \lambda^{2}_{Z_{c \bar b}}}  \,\, \rho_2(\rm{M^2},\rm{s_0}),\\
  \mu_{Z_{c \bar b}}^{J_\alpha^2}\,  &= \frac{e^{\frac{m_{Z_{c \bar b}}^{2}}{\rm{M^2}}}}{ \lambda^{2}_{Z_{c \bar b}}} \,\, \rho_3(\rm{M^2},\rm{s_0}),~~~~
 \mathcal{D}_{\mathrm{Z_{c \bar b}}}^{J_\alpha^2} = m_{\mathrm{Z_{c \bar b}}}^2 \frac{e^{\frac{m_{Z_{c \bar b}}^{2}}{\rm{M^2}}}}{ \lambda^{2}_{Z_{c \bar b}}}  \,\, \rho_4(\rm{M^2},\rm{s_0}),
 \end{align}

As the explicit forms of $\rho_i(\rm{M^2},\rm{s_0})$ are similar, for illustrative purposes, the result of $\rho_1(\rm{M^2},\rm{s_0})$ is provided below
\begin{align}
\rho_1(\rm{M^2},\rm{s_0})&= -\frac {1} {2 ^{19} \times 3^2 \times 5^2 \times 7  \pi^5}\Bigg[ 
   9 e_q \Big (-4 I[0, 6] + 
       42 m_c m_s \big (I[0, 5] + 15 I[1, 4]\big) - 
       129 I[1, 5]\Big) + 
    14  e_b \Big (38 m_c m_s I[0, 5] \nonumber\\
    &- 6 I[0, 6] + 
        130 m_c m_s I[1, 4] - 27 I[1, 5]\Big)\Bigg]\nonumber\\
       & +\frac {\langle g_s^2 G^2\rangle \langle \bar q q \rangle} {2 ^{21} \times 3^4  \pi^3}\Bigg[ 
   160 (e_c - e_s) m_b \big (I[0, 2] - 2 I[1, 1]\big) - 
    43 e_q m_c  (I_ 1[\mathcal S] + I_ 1[\mathcal {\tilde S}]) I[0, 
       2]\Bigg]\nonumber\\
       &-\frac {\langle g_s^2 G^2\rangle } {2 ^{22} \times 3^5 \times 5   \pi^5}\Bigg[ 5 (e_c - e_s) \big (21 I[0, 4] - 
       172 I[1, 3]\big) - 162 e_q \big (I[0, 4] - 4 I[1, 3]\big) - 
    12 e_b \big (3 I[0, 4] + 32 I[1, 3]\big)\Bigg]\nonumber\\
   & +\frac {\langle \bar q q \rangle } {2 ^{16} \times 3^3 \times 5   \pi^3}\Bigg[ 
   48 e_b m_b \big (20 m_c m_s I[0, 3] - 3 I[0, 4]\big) + 
    e_c m_c\Big (   (60 m_c m_s I[0, 3] - 
           9 I[0, 4]) (I_ 4[\mathcal S] - 
           I_ 4[\mathcal {\tilde S}]) \nonumber\\
           &- 
        4  (44 m_c m_s I[0, 3]- 9 I[0, 4]) I_ 5[h_\gamma]\Big)\Bigg]\nonumber\\
       & -\frac {\langle \bar ss \rangle } {2 ^{18} \times 3^3 \times 5   \pi^3}\Bigg[ 
   16 \Big (e_b m_c (-6 I[0, 4] + 66 m_ 0^2 I[1, 2] - 64 I[1, 3]) + 
       27 e_q m_c (I[0, 4] - 4 I[1, 3]) \nonumber\\
           &+ 
       9 e_b m_s (I[0, 4] - 4 I[1, 3])\Big) + 
    e_s m_c (I_ 2[\mathcal S] + 44 I_ 3[\mathcal S]) I[0, 4]\Bigg]\nonumber\\
    &+\frac {f_{3\gamma}} {2 ^{22} \times 3^3 \times 5^2 \times 7  \pi^3}\Bigg[
   112 e_q  (55 m_c m_s I[0, 4] - 9 I[0, 5]) I_ 1[\mathcal V] + 
    2511 e_s I_ 2[\mathcal V] I[0, 5]\Bigg],
        \label{app1}
\end{align}
where the $I[n,m]$, and~$I_i[\mathcal{F}]$ functions are expressed as:
\begin{align}
 I[n,m]&= \int_{\mathcal M}^{\rm{s_0}} ds ~ e^{-s/\rm{M^2}}~
 s^n\,(s-\mathcal M)^m,
 \end{align}
 \begin{align}
 I_1[\mathcal{F}]&=\int D_{\alpha_i} \int_0^1 dv~ \mathcal{F}(\alpha_{\bar q},\alpha_q,\alpha_g)
 \delta'(\alpha_ q +\bar v \alpha_g-u_0),\nonumber\\
  I_2[\mathcal{F}]&=\int D_{\alpha_i} \int_0^1 dv~ \mathcal{F}(\alpha_{\bar q},\alpha_q,\alpha_g)
 \delta'(\alpha_{\bar q}+ v \alpha_g-u_0),\nonumber\\
   I_3[\mathcal{F}]&=\int D_{\alpha_i} \int_0^1 dv~ \mathcal{F}(\alpha_{\bar q},\alpha_q,\alpha_g)
 \delta(\alpha_ q +\bar v \alpha_g-u_0),\nonumber\\
   I_4[\mathcal{F}]&=\int D_{\alpha_i} \int_0^1 dv~ \mathcal{F}(\alpha_{\bar q},\alpha_q,\alpha_g)
 \delta(\alpha_{\bar q}+ v \alpha_g-u_0),\nonumber\\
 I_5[\mathcal{F}]&=\int_0^1 du~ \mathcal{F}(u),\nonumber
 \end{align}
 where  $\mathcal M = (m_c+m_b+m_s)^2$  for the $[sc] [\bar q \bar b]$ states and $\mathcal M = (m_c+m_b+2m_s)^2$ for the $[sc] [\bar s \bar b]$ state; and $\mathcal{F}$ denotes the relevant DAs of the photon.

 It is important to note that the derivation of Eq. (\ref{edmn04})  is contingent upon the assumption that the hadron level of the QCD light-cone sum rules can be adequately approximated by a single pole. With regard to the tetra- or pentaquark states, it is essential to validate the aforementioned approximation through the introduction of additional arguments.   This is due to the fact that the hadron level of the QCD sum rules encompasses potential contributions from the two meson intermediate states (for details see for instance~\cite{Weinberg:2013cfa,Lucha:2021mwx,Kondo:2004cr,Lucha:2019pmp}).  Thus, it may be important to consider the contribution of two-meson intermediate states when attempting to extract the parameters related to tetra- or pentaquark states. Such contributions may be deducted from the QCD sum rules or incorporated into the parameters of the pole term. The first scheme was used in the study of pentaquarks~\cite{Sarac:2005fn,Wang:2019hyc,Lee:2004xk}, whereas the second scheme was utilized to extract tetraquarks~\cite{Wang:2015nwa,Agaev:2018vag,Sundu:2018nxt,Albuquerque:2021tqd,Albuquerque:2020hio,Wang:2020iqt,Wang:2019igl,Wang:2020cme}.
 In the context presented herein, it is of the utmost importance that the quark propagator be modified in a manner consistent with the following equation:
\begin{align}
\frac{1}{m^{2}-p^{2}} \longrightarrow  \frac{1}{m^{2}-p^{2}-i\sqrt{p^{2}}\,\Gamma (p)%
},  \label{eq:Modif}
\end{align}
where $\Gamma (p)$ is the finite width of the tetra- or pentaquark states generated by the intermediate two-meson contributions. When these contributions are properly incorporated into the QCD sum rules, it is demonstrated that they exert an influence on the physical observables to the extent of approximately $(5-7)\%$~(see Refs. \cite{Wang:2015nwa,Agaev:2018vag,Sundu:2018nxt,Albuquerque:2021tqd,Albuquerque:2020hio,Wang:2020iqt,Wang:2019igl,Wang:2020cme}), and that these contributions do not exceed the inherent errors associated with the QCD sum rule calculations. It is reasonable to posit that the results of the multipole moments will remain unaffected by the aforementioned contributions. Consequently, the contributions of the two-meson intermediate states at the hadron level of the correlation function can be safely ignored, and the zero-width single-pole scheme can be used instead.

Analytical expressions for the multipole moments of the $[sc] [\bar q \bar b]$ and $[sc] [\bar s \bar b]$ tetraquarks with $J^P = 1^+$  have been derived. The subsequent section will present a numerical analysis of these quantities.
\end{widetext}

\section{Numerical analysis of the multipole moments}\label{numerical}

To conduct numerical calculations of QCD sum rules for multipole moments, several input quantities are required, as given in Table~\ref{inputparameter}.  In numerical analysis, we set $m_u$ =$m_d$ = 0 and $m^2_s = 0$, but consider terms proportional to $m_s$.  Another crucial input parameter in the numerical analysis is the photon DAs and the wave functions employed therein. These expressions and the input parameters used in their explicit forms are taken from Ref.~\cite{Ball:2002ps}.
%
  \begin{table}[htp]
	\addtolength{\tabcolsep}{10pt}
	\caption{Parameters employed as inputs in the calculations~\cite{Workman:2022ynf,Ioffe:2005ym,Narison:2018nbv,Agaev:2017uky,Wang:2020jgb}.}
	\label{inputparameter}
\begin{tabular}{l|c|ccccc}
               \hline\hline
Parameter & Value&Unit \\
                                        \hline\hline
                                        %
$m_s$&$ 93.4^{+8.6}_{-3.4}$&MeV         \\
$m_c$&$ 1.67 \pm 0.07$&GeV                 \\
$m_b$&$ 4.78 \pm 0.06$&GeV                    \\
$m_{[sc] [\bar s \bar b]}$&$  7.30 \pm 0.77 $&GeV                       \\
$m_{[sc] [\bar q \bar b]}$&$  7.18 \pm 0.76 $&GeV    \\ 
$m_0^{2} $&$ 0.8 \pm 0.1 $&\,\,GeV$^2$                        \\
$f_{3\gamma} $&$ -0.0039 $&\,\,GeV$^2$     \\
$\langle \bar qq\rangle $&$ (-0.24 \pm 0.01)^3 $&\,\,GeV$^3$                    \\
$\langle \bar ss\rangle $&$ 0.8 $    $\langle \bar qq\rangle $ &\,\,GeV$^3$               \\
$ \langle g_s^2G^2\rangle  $&$ 0.48 \pm 0.14 $&\,\,GeV$^4$                        \\
$\lambda_{[sc] [\bar q \bar b]}$&$  0.035 \pm 0.011    $&\,\,GeV$^5$                        \\
$\lambda_{[sc] [\bar s \bar b]}$&$  0.046 \pm 0.014   $&\,\,GeV$^5$                                                                  \\
                                      \hline\hline
 \end{tabular}
\end{table}

Besides the aforementioned input parameters, two further parameters are required for the calculations: the continuum threshold parameter $\rm{s_0}$ and the Borel mass $\rm{M^2}$. To acquire reliable results from QCD sum rules, the dependence of the multipole moments on $\rm{s_0}$ and $\rm{M^2}$ should be relatively weak, which is called working windows. The working windows of these parameters are determined from the standard procedures of the method used. These procedures are known as  pole dominance (PC) and convergence of the OPE (CVG), which are described by the following formulas: 
\begin{align}
 \mbox{PC} &=\frac{\rho_i (\rm{M^2},\rm{s_0})}{\rho_i (\rm{M^2},\infty)},
 \\
 \nonumber\\
 \mbox{CVG} (\rm{M^2}, \rm{s_0}) &=\frac{\rho_i^{\rm{Dim 7}} (\rm{M^2},\rm{s_0})}{\rho_i (\rm{M^2},\rm{s_0})},
 \end{align}
 where $\rho_i^{\rm{Dim 7}} (\rm{M^2},\rm{s_0})$ stands for the highest dimensional term in the OPE  of $\rho_i (\rm{M^2},\rm{s_0})$. According to the sum rules analysis, the CVG must be sufficiently small to guarantee OPE convergence, while the PC must be sufficiently large to maximize the efficiency of the single-pole scheme. Under the aforementioned prerequisites, one can obtain the working windows of the $\rm{M^2}$ and $\rm{s_0}$, which are presented in Table \ref{table}. As a result, in the same table, the values achieved for the PC and CVG are also listed. For completeness, Figs. \ref{figMsq} and \ref{figs0} illustrate the variations of the extracted multipole moments of these states concerning $\rm{M^2}$ and $\rm{s_0}$. As illustrated by the presented figures, the multipole moments of these states show a relatively mild dependence on the parameter $\rm{M^2}$. Although the multipole moments of these states exhibit some dependence on $\rm{s_0}$, they remain within the limits permitted by this approach and constitute the primary source of uncertainty.
%
%
  \begin{table}[htb!]
	\addtolength{\tabcolsep}{10pt}
	\caption{Predicted multipole moments of the $[sc] [\bar q \bar b]$ and $[sc] [\bar s \bar b]$ states with $J^P = 1^+$.}
	\label{table}
	\begin{center}
\begin{tabular}{lccccccc}
	   \hline\hline
	   \\
	   Tetraquarks &  $\mu$\,\,[$\rm{\mu_N}$]	& $\mathcal D (\times 10^{-2})$\,\,[$\rm{fm^2}$]& $\rm{M^2}\,\,[\rm{GeV}^2]$& $\rm{s_0}\,\,[\rm{GeV}^2]$& PC\,\,[$\%$] & CVG\,\,[$\%$]	   \\
	   \\
	   \hline\hline
	  \\
	   $[sc] [\bar u \bar b]$&   $ -2.12^{+0.74}_{-0.59}$     &$-0.66^{+0.15}_{-0.13}$ & [4.0, 4.6]& [59, 62]& [53.3, 29.4]& $< 2.0$\\
	   \\
	   \\
	 $[sc] [\bar d \bar b]$&   $ 1.66^{+0.60}_{-0.46}$     &$0.33^{+0.07}_{-0.07}$ & [4.0, 4.6]& [59, 62]& [52.8, 29.1]& $< 2.0$\\
	   \\
	   \\
	    $[sc] [\bar s \bar b]$&   $ 2.01^{+0.61}_{-0.50}$     &$0.37^{+0.07}_{-0.07}$ & [4.4, 5.0]& [61, 64]& [51.7, 28.3]& $< 2.0$\\
	    \\
	   \hline\hline
\end{tabular}
\end{center}
\end{table}
%
 Predicted multipole moments of the $[sc] [\bar q \bar b]$ and $[sc] [\bar s \bar b]$ states with $J^P = 1^+$, which consider the uncertainties inherent in the input parameters and the variation in the $\rm{M^2}$ and $\rm{s_0}$ working windows, are presented in Table~\ref{table}.  
In light of the findings presented in this study, the following key points have been identified:
  \begin{itemize}
    
\item Our analysis shows that the multipole moments of these states are governed by axial-vector diquark ([$\bar q \bar b$]/[$\bar s \bar b$]) 
 component of the interpolating currents.
 
\item When we analyze the individual quark contributions (which can be achieved through the charge factors $e_q$, $e_c$, and $e_b$), it becomes evident that the magnetic dipole moments are primarily influenced by light quarks, which account for approximately $74\%$ of the total, while heavy quarks represent the remaining $26\%$.  In the case of the quadrupole moments, the total contributions come from the light quarks. Upon closer examination of the quadrupole moments, it becomes evident that the missing heavy quark contributions are due to the terms containing heavy quarks exactly canceling each other out.

\item The order of the magnetic dipole moments could be used to give an idea of the experimental accessibility of the corresponding physical quantities. Their order of magnitude would suggest that these outcomes may be achievable in forthcoming experiments.
  
\item The quadrupole moment results obtained for these states are non-zero, indicating the presence of a non-spherical charge distribution. By the predicted signs, the geometric shape of the $[sc] [\bar u \bar b]$ state is oblate, whereas the $[sc] [\bar d \bar b]$ and $[sc] [\bar s \bar b]$ states are prolate.
  
\item $U$-symmetry violation between the $[sc] [\bar d \bar b]$ and $[sc] [\bar s \bar b]$ states in case of the magnetic dipole moment is about $ 18\%$, while in the case of the quadrupole moment, it is roughly $ 12\%$. A reasonable $U$-symmetry violation is observed.
  
\end{itemize}

\section{Concluding remarks}\label{sum}

A systematic study of the electromagnetic properties of exotic states is conducted to elucidate their nature, which continues to be the subject of controversy and incomplete understanding in the field. This work estimates the multipole moments of possible $[sc] [\bar q \bar b]$ and $[sc] [\bar s \bar b]$ states within the context of the QCD light-cone sum rules. In calculating the multipole moments of the $[sc] [\bar q \bar b]$ and $[sc] [\bar s \bar b]$ states, the diquark-antidiquark configuration is taken into account with $J^P = 1^+$ spin-parity quantum numbers. The order of magnitude of the magnetic dipole moment demonstrates that it may  be measured in an experiment. The predictions derived in this study can be checked through the use of other theoretical models, such as lattice QCD and chiral perturbation theory. The multipole moments of the $[sc] [\bar q \bar b]$ and $[sc] [\bar s \bar b]$ states reveal valuable knowledge concerning the internal structure, size, and shape of the hadrons.  The determination of these parameters could be a significant step in our interpretation of hadron properties concerning quark-gluon degrees of freedom and in elucidating their nature.  It is similarly crucial to ascertain the branching ratios of the various decay modes/channels of these tetraquark states. It would be highly intriguing to anticipate forthcoming experimental efforts that will investigate the potential existence of $[sc] [\bar q \bar b]$ and $[sc] [\bar s \bar b]$ states. These findings would then be subjected to rigorous testing to validate and expand upon the present results. To gain a more comprehensive understanding of these findings, further research is recommended.

 \begin{widetext}
 
 \begin{figure}[htp]
\centering
  \subfloat[]{\includegraphics[width=0.45\textwidth]{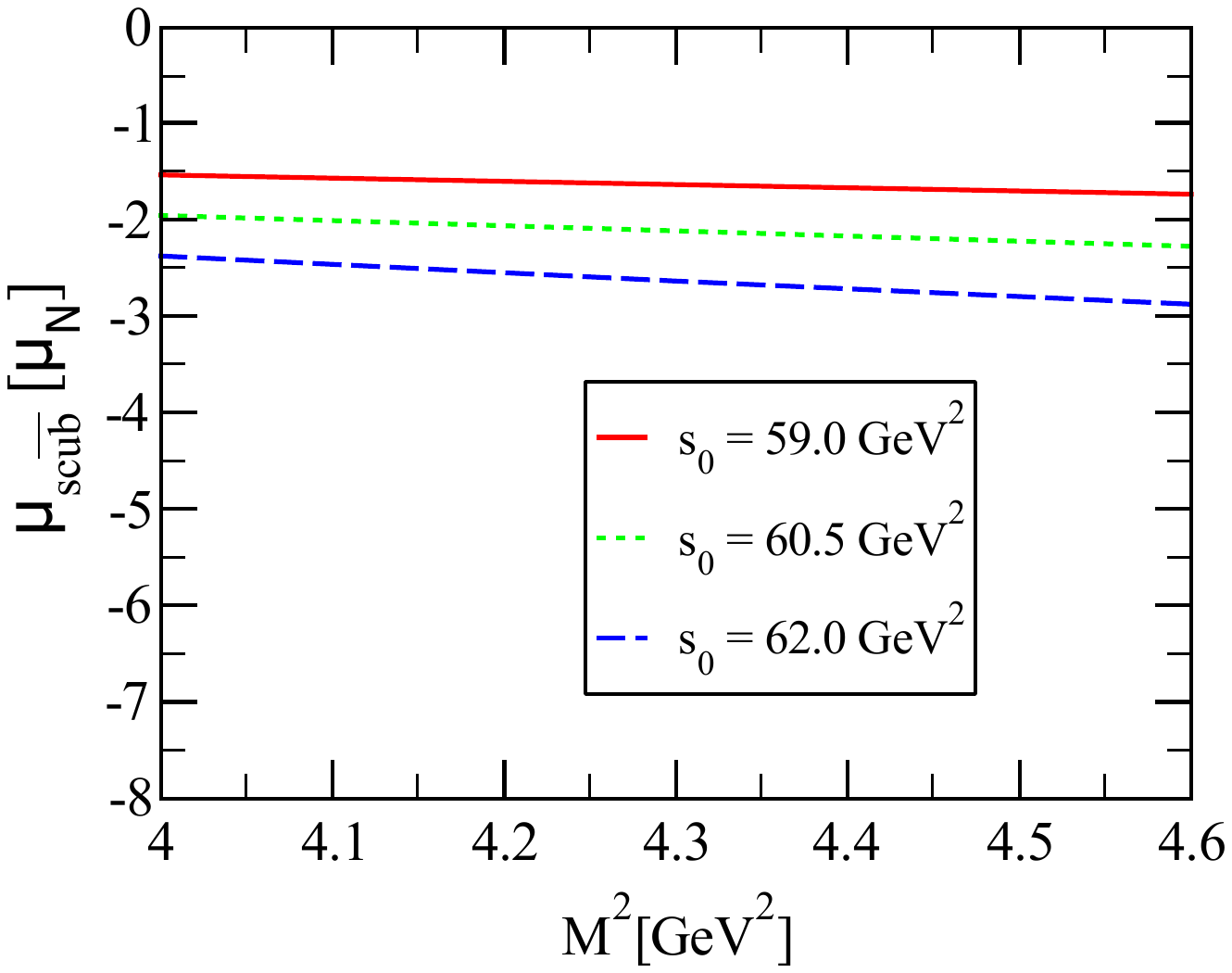}} ~~~
  \subfloat[]{\includegraphics[width=0.45\textwidth]{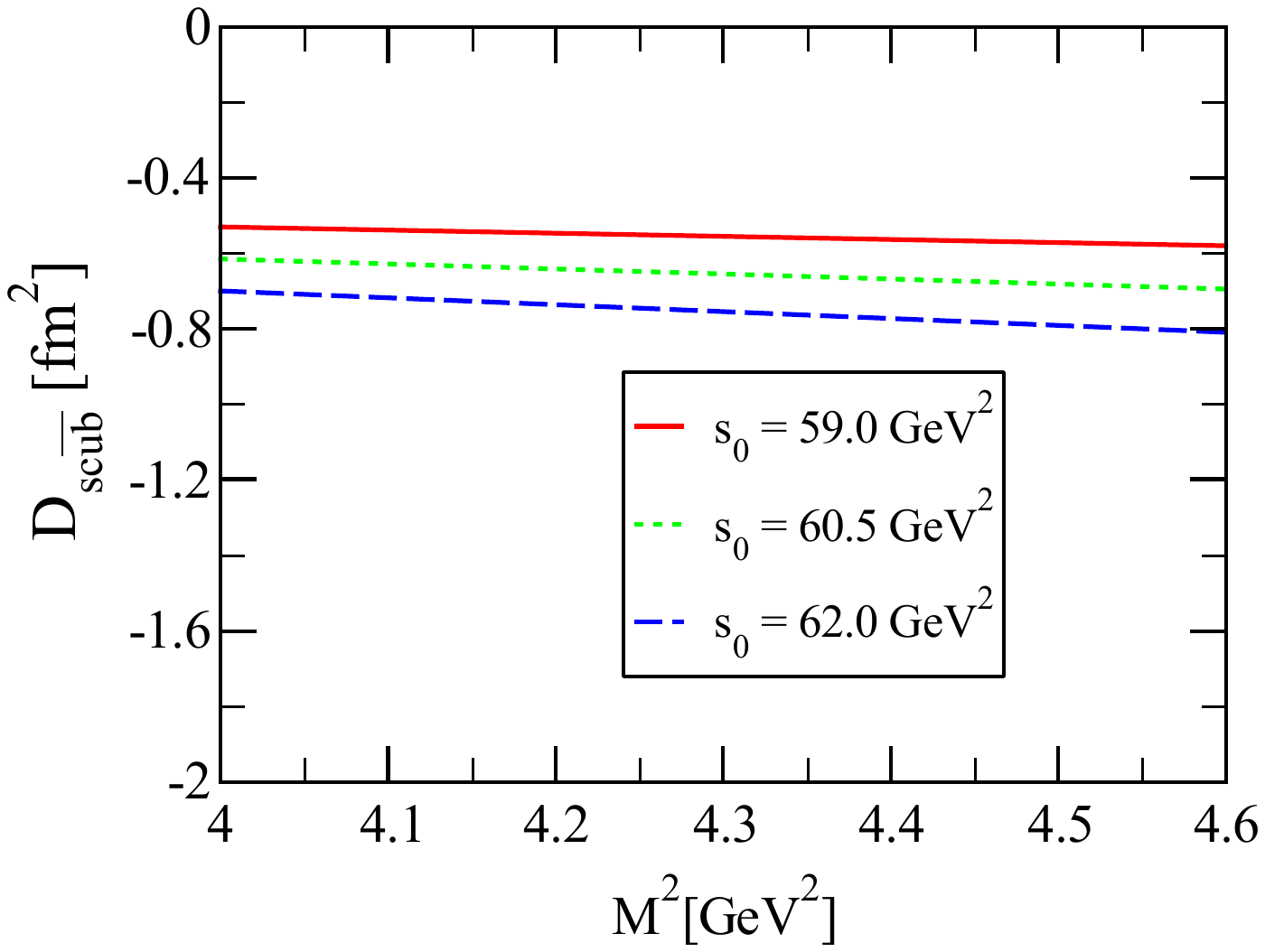}} \\
  \subfloat[]{\includegraphics[width=0.45\textwidth]{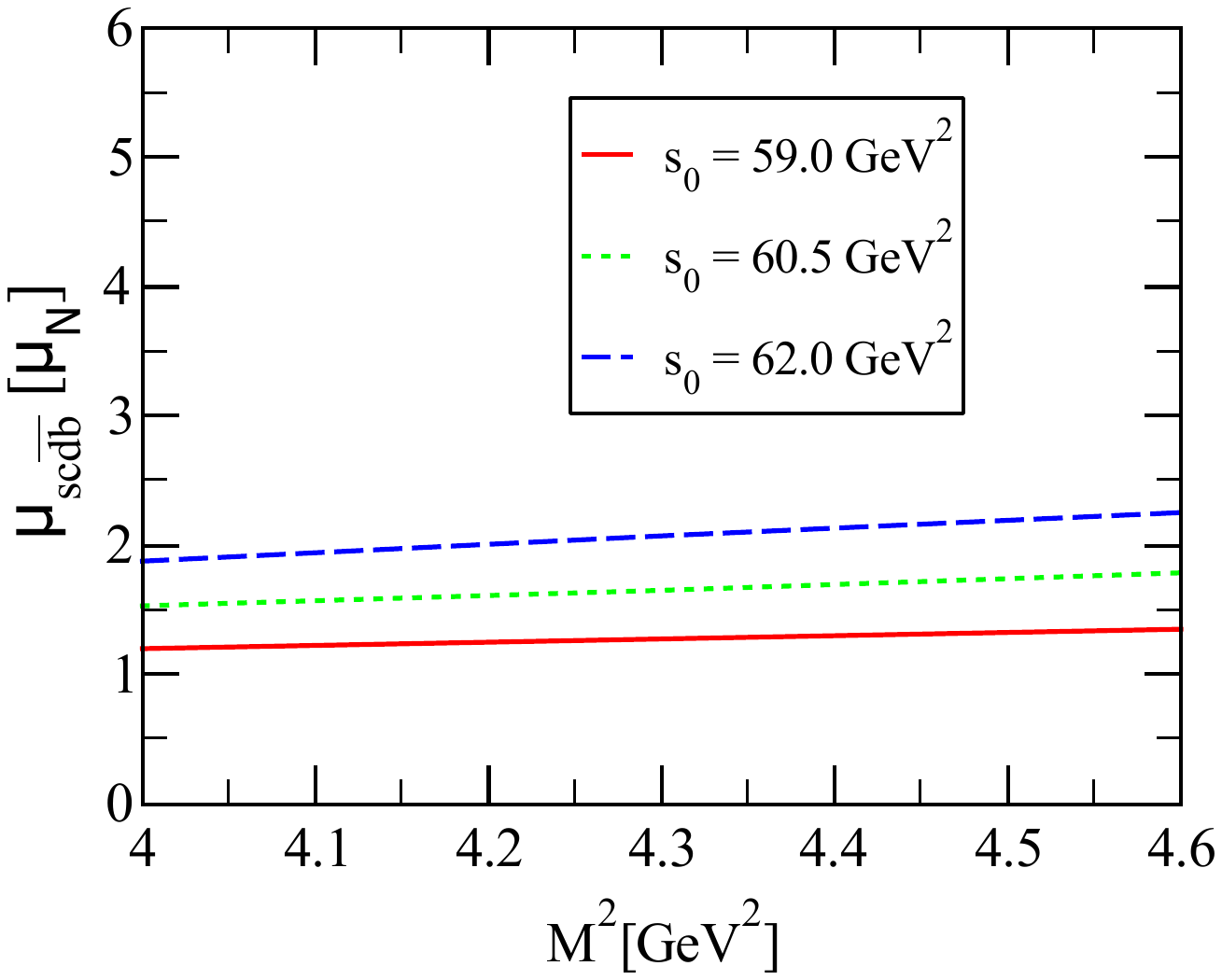}} ~~~
  \subfloat[]{\includegraphics[width=0.45\textwidth]{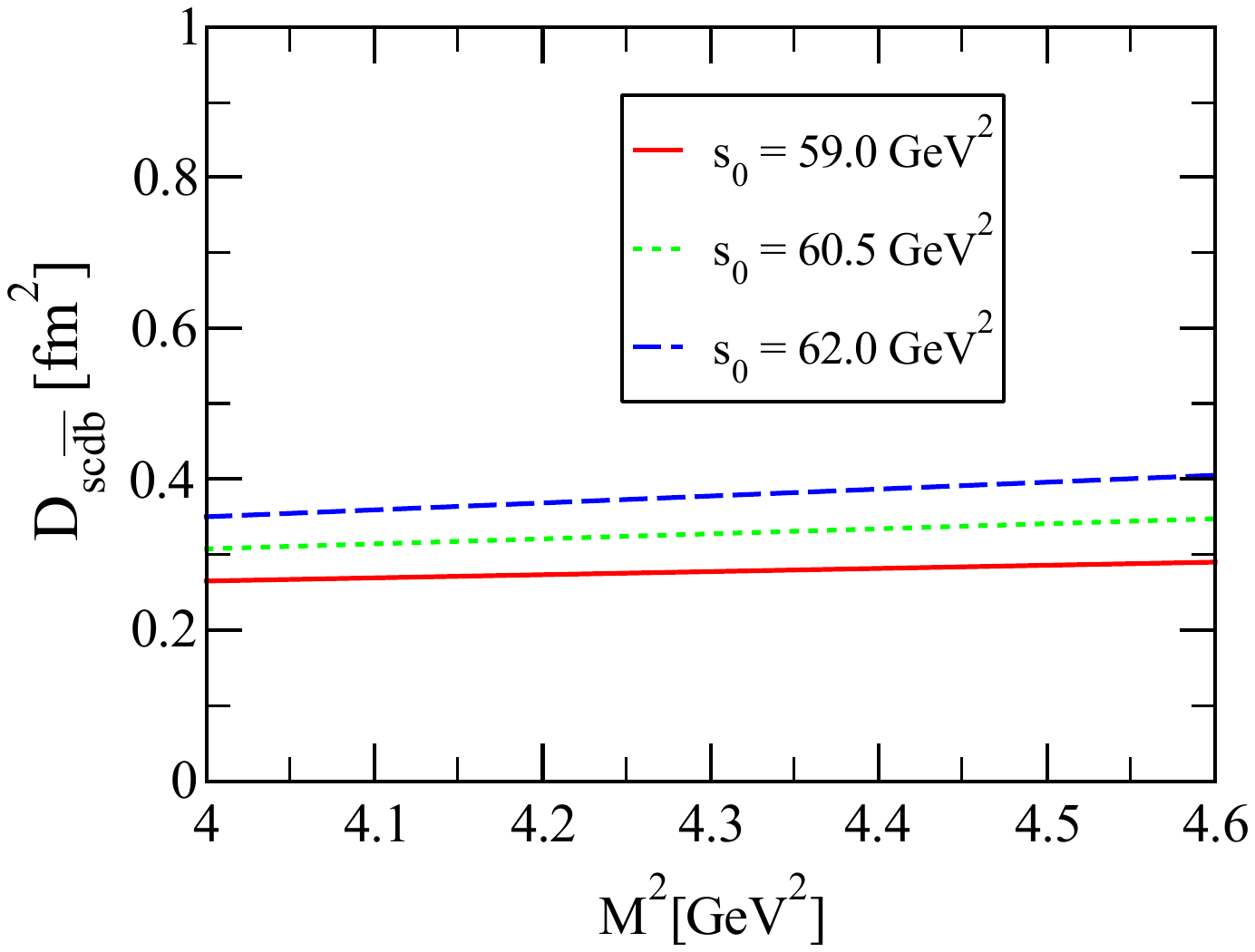}}\\
  \subfloat[]{\includegraphics[width=0.45\textwidth]{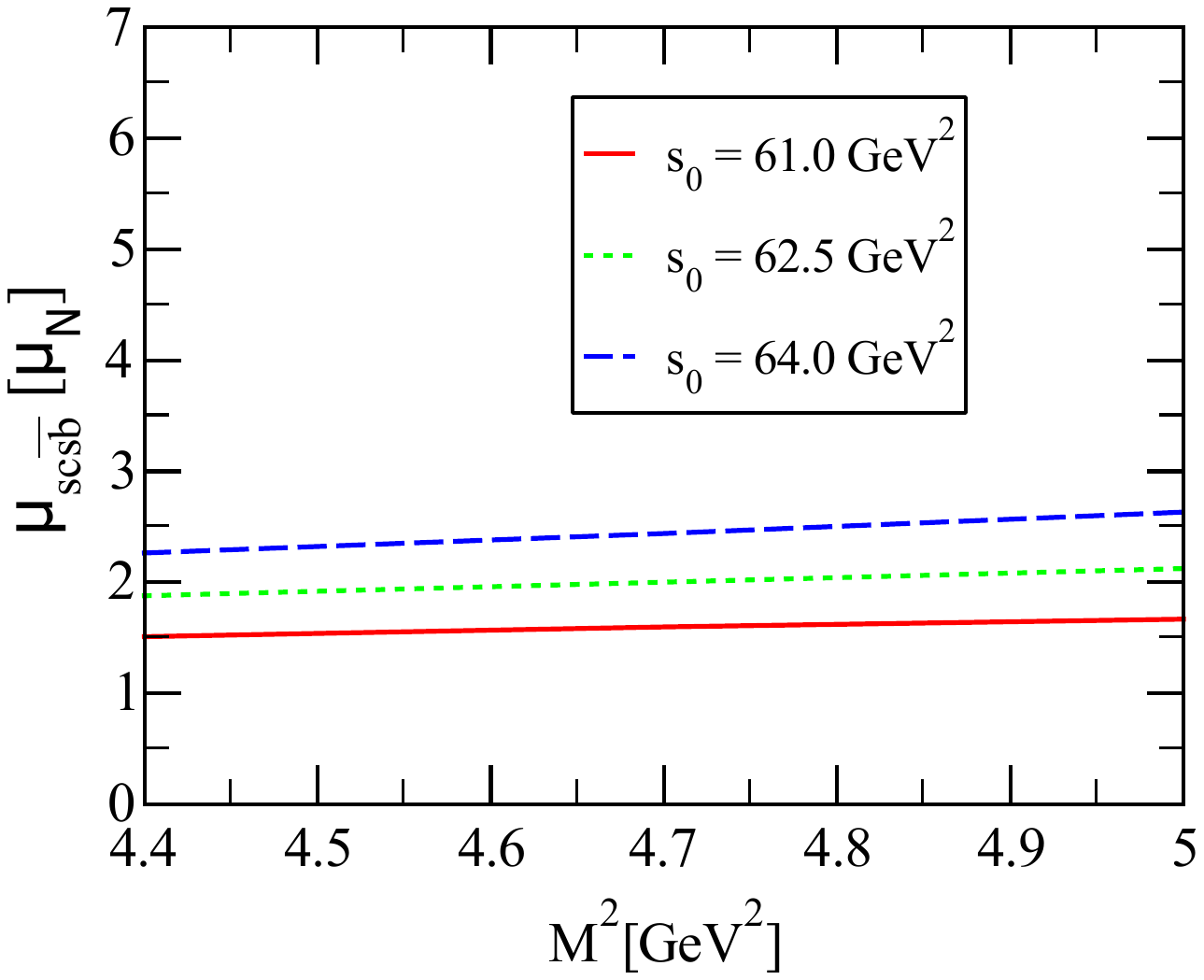}} ~~~
  \subfloat[]{\includegraphics[width=0.45\textwidth]{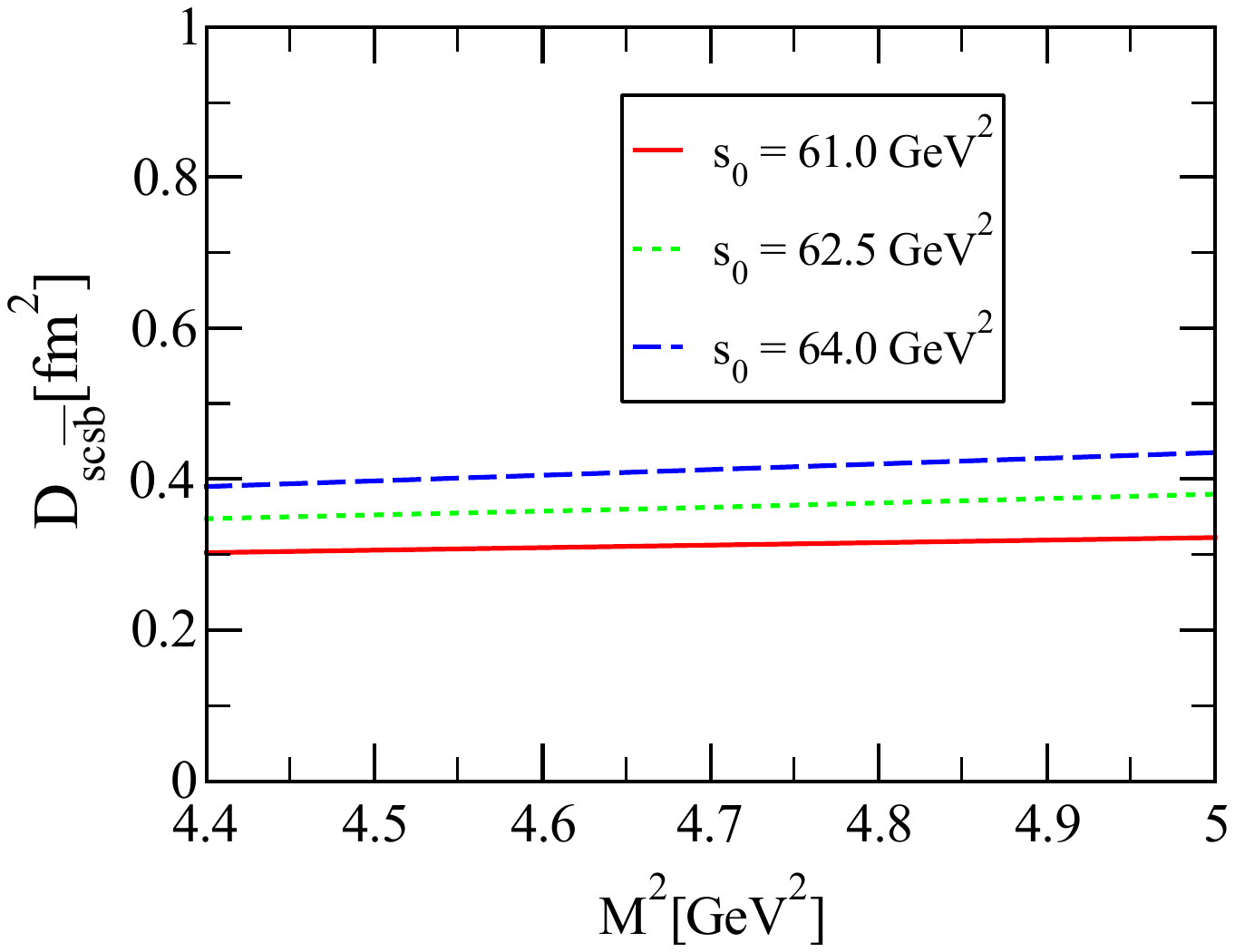}}
  \caption{ Variation of multipole moments $[sc] [\bar q \bar b]$ and $[sc] [\bar s \bar b]$ states as a function of the $\rm{M^2}$ at different values of $\rm{s_0}$.}
 \label{figMsq}
  \end{figure}

 \begin{figure}[htp]
\centering
  \subfloat[]{\includegraphics[width=0.45\textwidth]{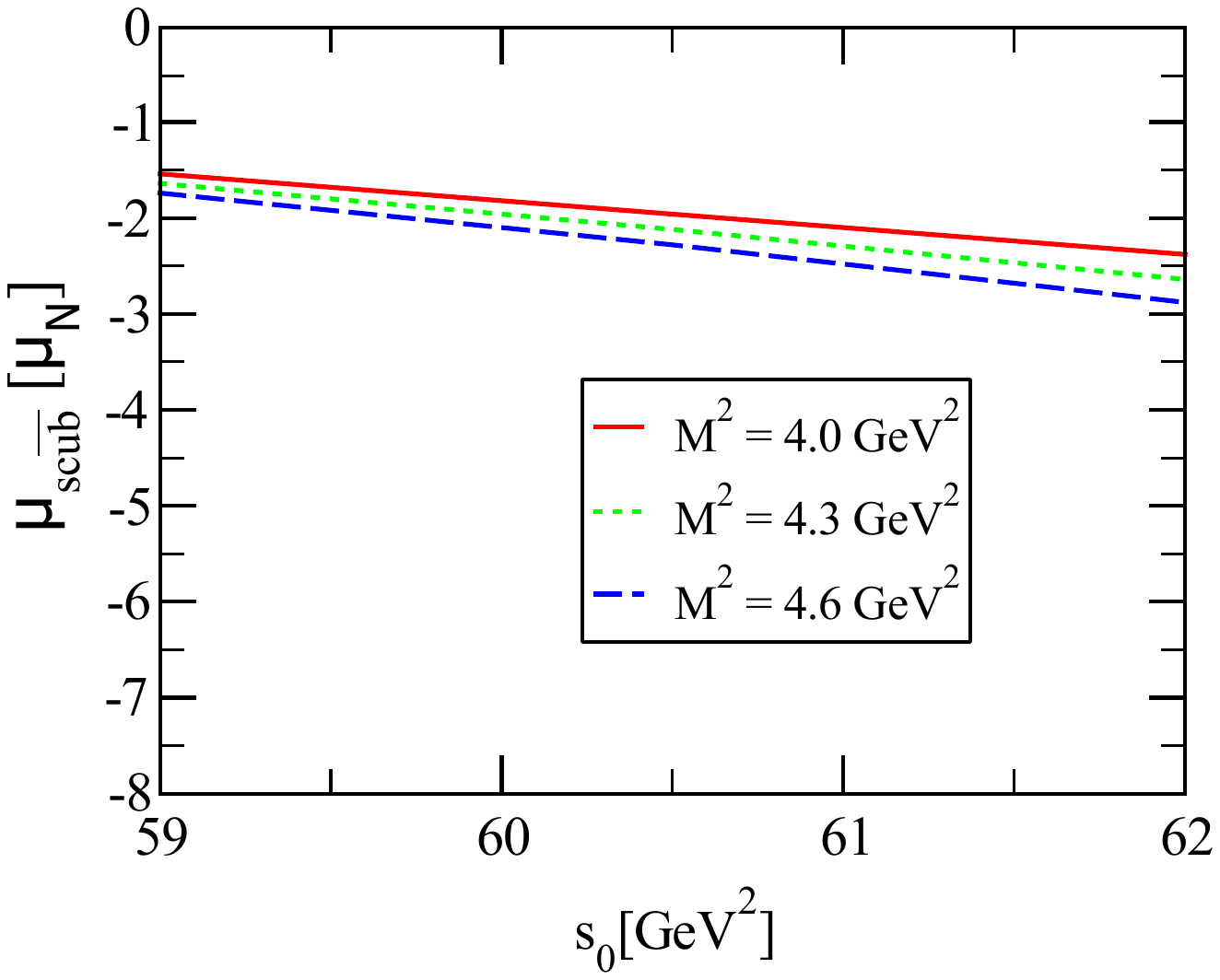}} ~~~
  \subfloat[]{\includegraphics[width=0.45\textwidth]{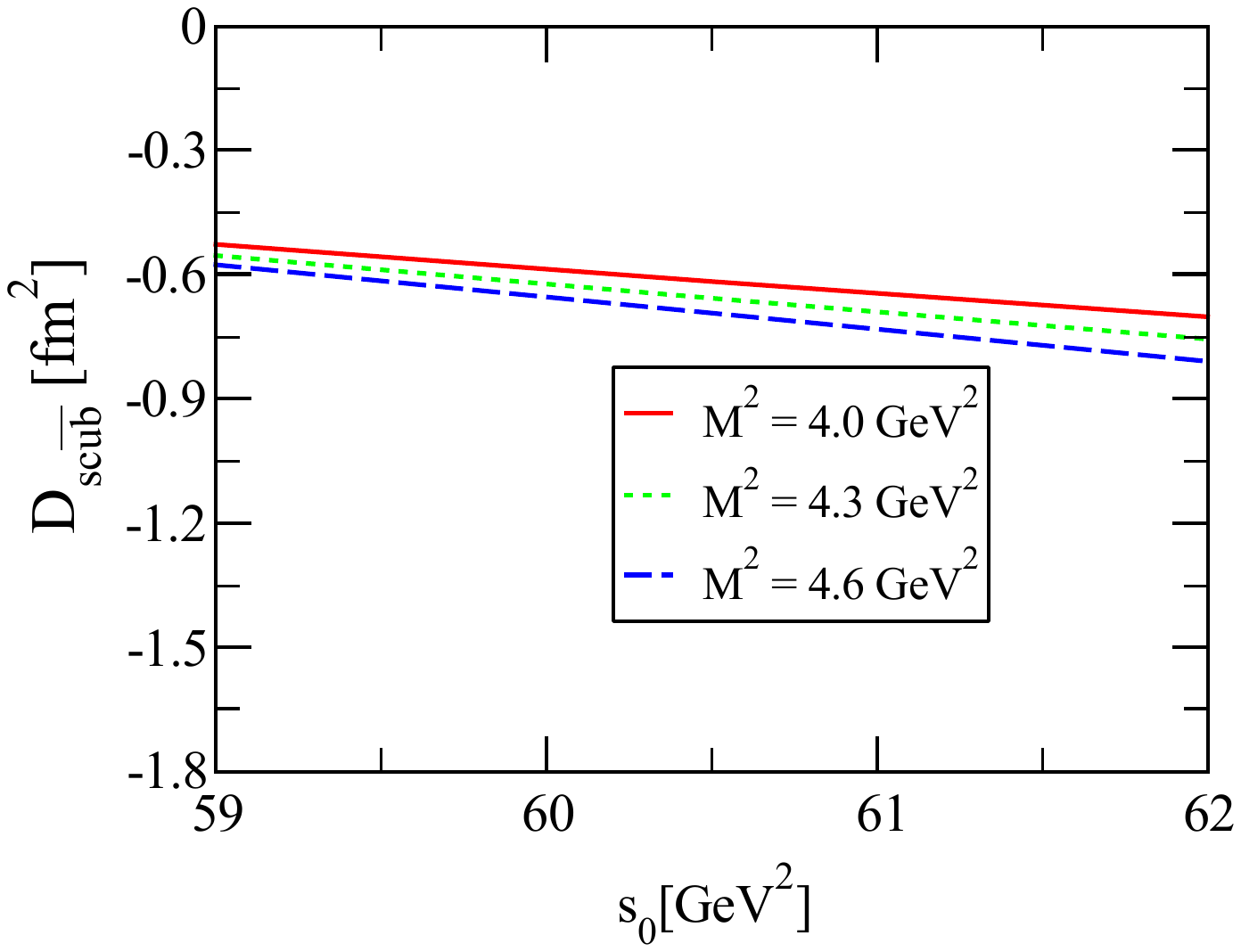}} \\
  \subfloat[]{\includegraphics[width=0.45\textwidth]{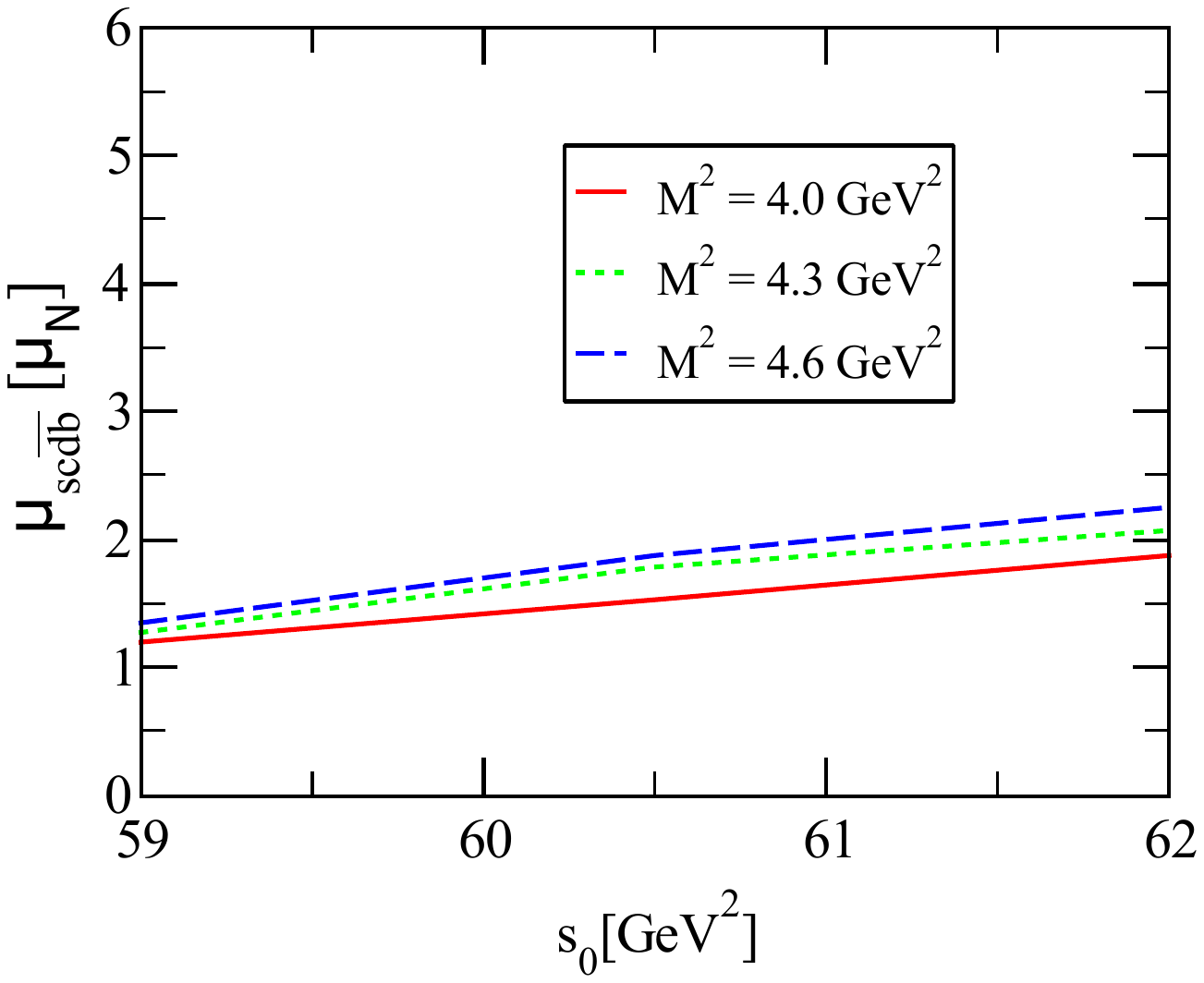}} ~~~
  \subfloat[]{\includegraphics[width=0.45\textwidth]{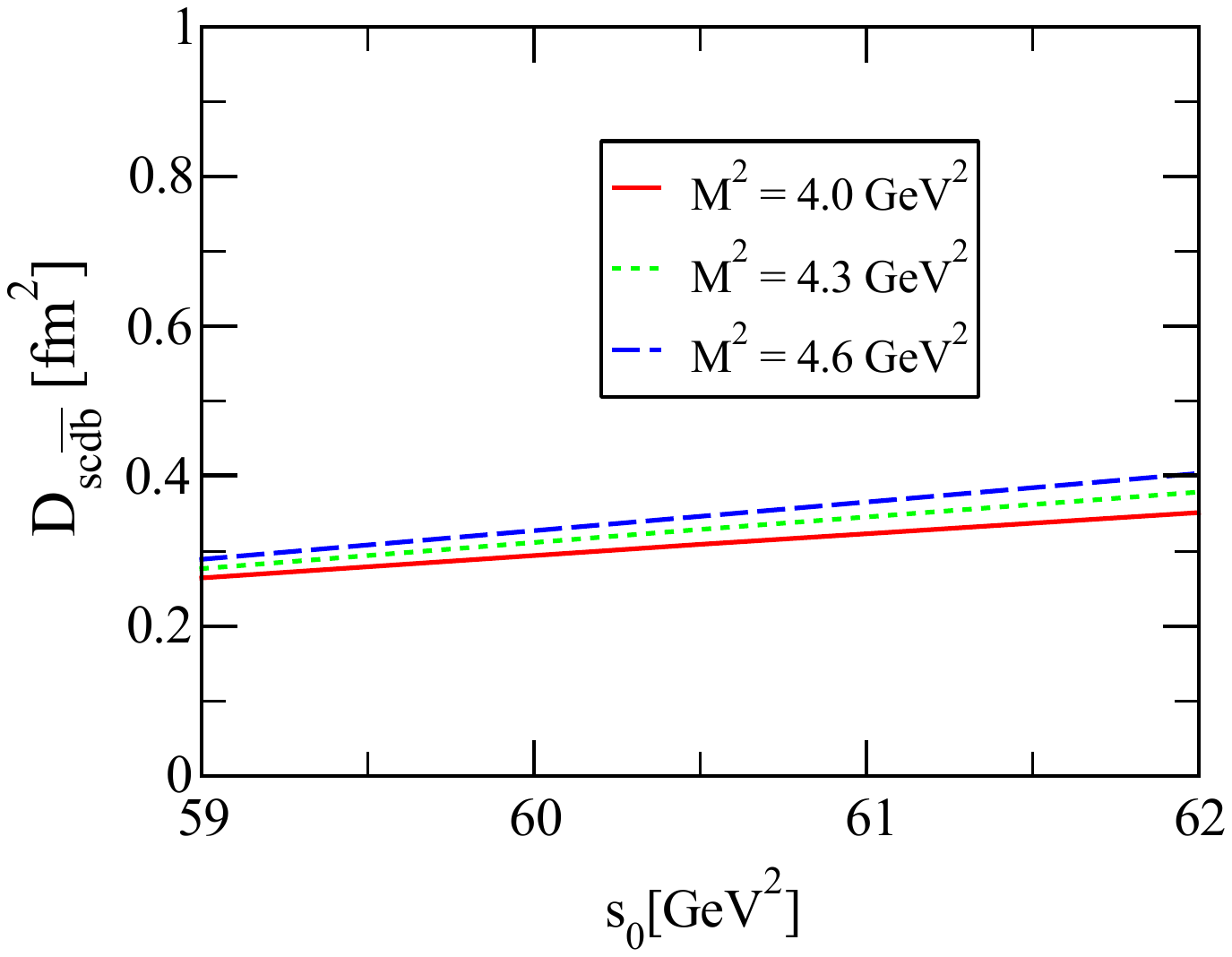}}\\
  \subfloat[]{\includegraphics[width=0.45\textwidth]{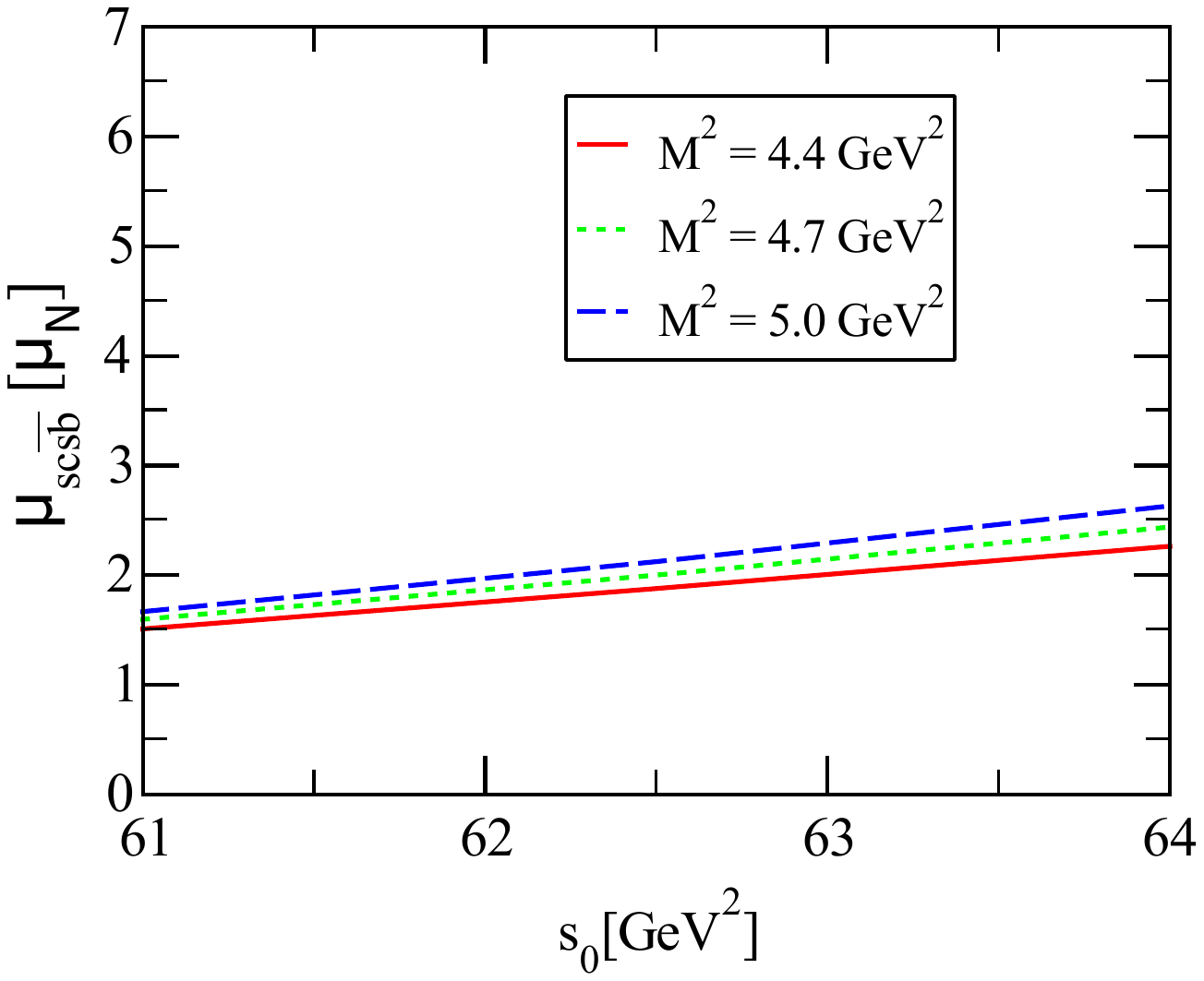}} ~~~
  \subfloat[]{\includegraphics[width=0.45\textwidth]{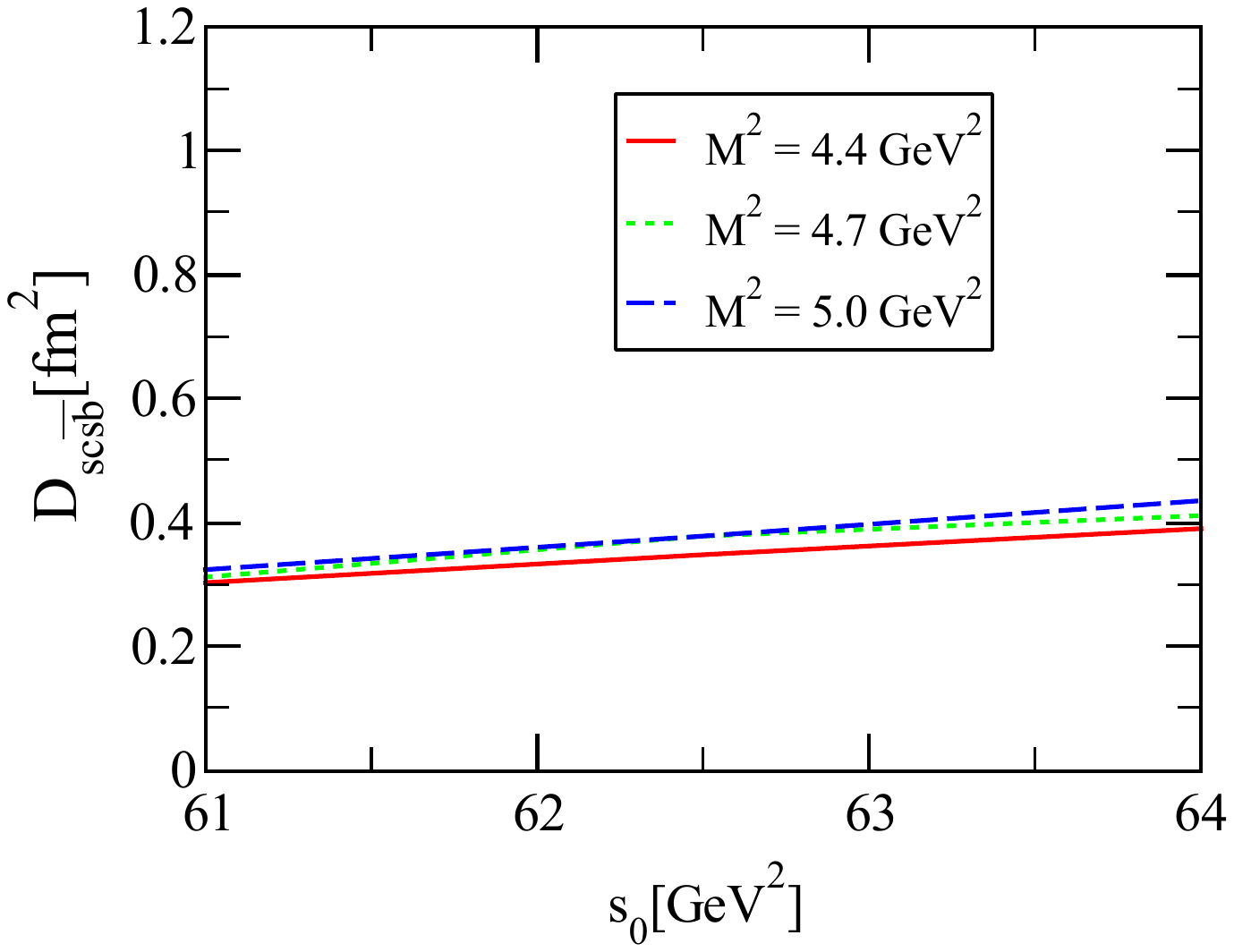}}
  \caption{ Variation of multipole moments $[sc] [\bar q \bar b]$ and $[sc] [\bar s \bar b]$ states as a function of the $\rm{s_0}$ at different values of $\rm{M^2}$.}
 \label{figs0}
  \end{figure}
  
   \end{widetext}

\bibliography{ZcbbarEM.bib}
\bibliographystyle{elsarticle-num}

\end{document}